\documentclass[letterpaper, 10 pt, conference]{ieeeconf}

\IEEEoverridecommandlockouts

\usepackage{graphicx}
\usepackage{amsmath}  
\usepackage{amssymb}  
\usepackage{bm}         
\usepackage{cite}     
\usepackage{float} 
\usepackage{array}
\usepackage{url}
\usepackage{subcaption}

 \usepackage{booktabs}

\title{\LARGE \bf A Multi-Objective Optimisation Framework for Corticomuscular EEG-EMG Pair Selection in Hybrid BCI}

\author{Dekka Muni Kumar$^1$ and Yogesh Kumar Meena$^1$ %
\thanks{$^{1}$Dekka Muni kumar and Yogesh Kumar Meena are with Human-AI Interaction
(HAIx) Lab, IIT Gandhinagar, India.
        {\tt\small yk.meena@iitgn.ac.in}}}

\begin{document}

\maketitle
\thispagestyle{empty}
\pagestyle{empty}
\raggedbottom

\begin{abstract}
Hybrid brain-computer interface (BCI) systems that integrate electroencephalography (EEG) and electromyography (EMG) signals have shown significant potential in improving the reliability of motor imagery (MI) classification, particularly in neuro-rehabilitation applications. However, identifying informative EEG-EMG channel pairs that effectively capture corticomuscular interactions remains a challenging problem, as existing approaches typically rely on manually predefined channel combinations that may not generalise across subjects. In this work, a data-driven EEG-EMG pair selection framework is proposed, in which channel pair selection is formulated as a constrained bi-objective optimisation problem. The proposed method jointly maximises the spatial relevance of EEG channels with respect to motor cortex regions and the corticomuscular coupling strength between EEG and EMG signals, and is solved using the NSGA-II to automatically identify an optimal subset of pairs. To extract discriminative features, the correlation between band-power time features capturing EEG-EMG interaction is combined with ERD-based EEG features, and a sliding-window-based temporal analysis is employed to account for the dynamic nature of MI signals. The proposed framework is evaluated on MI data from eight stroke patients and achieves an average classification accuracy of 89.6\%, demonstrating its effectiveness in capturing physiologically meaningful corticomuscular interactions and improving classification performance.
\end{abstract}


\section{Introduction}

Brain-computer interfaces (BCIs) have emerged as a promising paradigm for assistive technologies and neuro-rehabilitation, enabling direct communication between the human brain and external devices by translating neural activity into actionable control signals~\cite{wolpaw2002brain}. These systems have been widely explored for applications such as neuro-prosthetic control, wheelchair navigation, and upper-limb rehabilitation, particularly for patients suffering from motor impairments~\cite{muller2015brain, chowdhury2018eeg}. Among various BCI paradigms, motor imagery (MI)-based BCIs have gained significant attention due to their ability to decode imagined movements without requiring physical execution~\cite{pfurtscheller2001motor}. During MI tasks, characteristic modulations in sensorimotor rhythms (SMR) are observed, especially in the $\mu$ (8-13 Hz) and $\beta$ (13-30 Hz) frequency bands, which are commonly exploited for detecting motor intention~\cite{pfurtscheller1999event, blankertz2008optimizing}.

Despite the significant progress achieved by EEG-based BCIs, their practical deployment remains challenging due to the inherently low signal-to-noise ratio, susceptibility to physiological and environmental artefacts, and considerable inter- and intra-subject variability of neural signals, often necessitating extensive subject-specific calibration~\cite{lotte2018review}. In contrast, electromyography (EMG)-based systems directly measure muscle activity and are effective for detecting executed movements; however, their performance degrades in individuals with severe neuromuscular impairments and is influenced by muscle fatigue, spasticity, and trial-to-trial variability~\cite{liu2013emg,makowski2015emg}. These complementary limitations have motivated the development of hybrid BCIs that integrate multiple physiological modalities to achieve more robust and reliable motor intention decoding.

Hybrid BCIs improve the limitations of single-modality systems by integrating complementary control modalities. For example, gaze and motor imagery have been combined to improve single-trial detection performance and increase the number of available control commands~\cite{meena2015simultaneous,7318410}. Among hybrid paradigms, EEG-EMG BCIs have attracted considerable attention by jointly exploiting cortical and muscular activity~\cite{allison2011hybrid,muller2011hybrid}. Recent multimodal BCI studies have further demonstrated the benefits of integrating EEG with complementary physiological signals while highlighting adaptive multimodal feature selection and physiologically interpretable fusion as important future directions~\cite{Lee2025HybridEEG}. A key concept in EEG-EMG BCIs is corticomuscular interaction, which reflects the functional coupling between brain and muscle activity during motor tasks~\cite{conway1995synchronization,cincotti2012neurofeedback}.


Traditionally, corticomuscular interaction has been quantified using corticomuscular coherence (CMC), which measures the synchronisation between EEG and EMG signals in the frequency domain~\cite{halliday1998coherence}. Although CMC provides valuable insights into functional brain--muscle connectivity, it requires relatively long signal segments to achieve adequate frequency resolution and is sensitive to noise and the non-stationary characteristics of EEG signals, limiting its applicability to single-trial and real-time BCI systems~\cite{pfurtscheller2003cmc,tuncel2010cmc}. To address these limitations, Chowdhury \textit{et al.} proposed the Correlation between Band-Power Time courses (CBPT), which estimates corticomuscular interaction in the time domain by computing the correlation between the band-limited power envelopes of EEG and EMG signals~\cite{chowdhury2018eeg}. By exploiting the temporal relationship between EEG event-related desynchronization (ERD) and the corresponding EMG activation, CBPT enables reliable single-trial estimation of corticomuscular interaction and has demonstrated promising performance in hybrid EEG-EMG BCI systems for neurorehabilitation.
optimisation-based channel selection, and corticomuscular interaction analysis. 
Despite its effectiveness, existing CBPT-based frameworks employ manually predefined EEG-EMG channel pairs based on prior neurophysiological knowledge. Such fixed channel combinations do not account for inter-subject variability in cortical activation patterns or differences in corticomuscular connectivity, thereby limiting their adaptability and generalizability across users. 

Channel selection has been extensively investigated in EEG-based BCI systems to reduce redundant channels and improve classification performance. Existing approaches include filter-based, wrapper-based, and optimisation-driven techniques~\cite{baig2019survey,faye2022review}. In particular, evolutionary and swarm-based optimisation algorithms have been successfully employed to identify compact EEG channel subsets by optimising objectives such as classification accuracy and channel sparsity. However, these methods have been developed primarily for EEG-only BCIs and do not exploit the complementary information available from EMG signals. Consequently, the problem of selecting optimal EEG-EMG channel pairs in hybrid BCIs remains largely unexplored.

Motivated by these challenges, this work proposes a data-driven framework for selecting EEG-EMG pairs for hybrid BCI systems.  The proposed framework is evaluated on a dataset recorded in the previous research~\cite{chowdhury2018eeg}. The evaluation results demonstrate that the proposed framework improves classification accuracy by selecting meaningful EEG-EMG pairs. We provide three contributions in this paper, by:

\begin{itemize}
    \item Proposing a framework for selecting EEG-EMG channel pairs, where the selection is formulated as a constrained bi-objective optimisation problem rather than relying on manually predefined pairs.
    \item Developing a multi-objective optimisation framework that combines a Gaussian kernel-based spatial relevance measure with a corticomuscular coupling objective derived from CBPT features, resolved using the NSGA-II algorithm to automatically identify an optimal subset of EEG--EMG pairs.
    \item Presenting a hybrid feature extraction and classification framework that integrates CBPT and ERD features with sliding-window-based temporal selection and SVM classification, enabling robust MI detection.
\end{itemize}


\begin{figure}[!t]
    \centering
    \includegraphics[width=\linewidth]{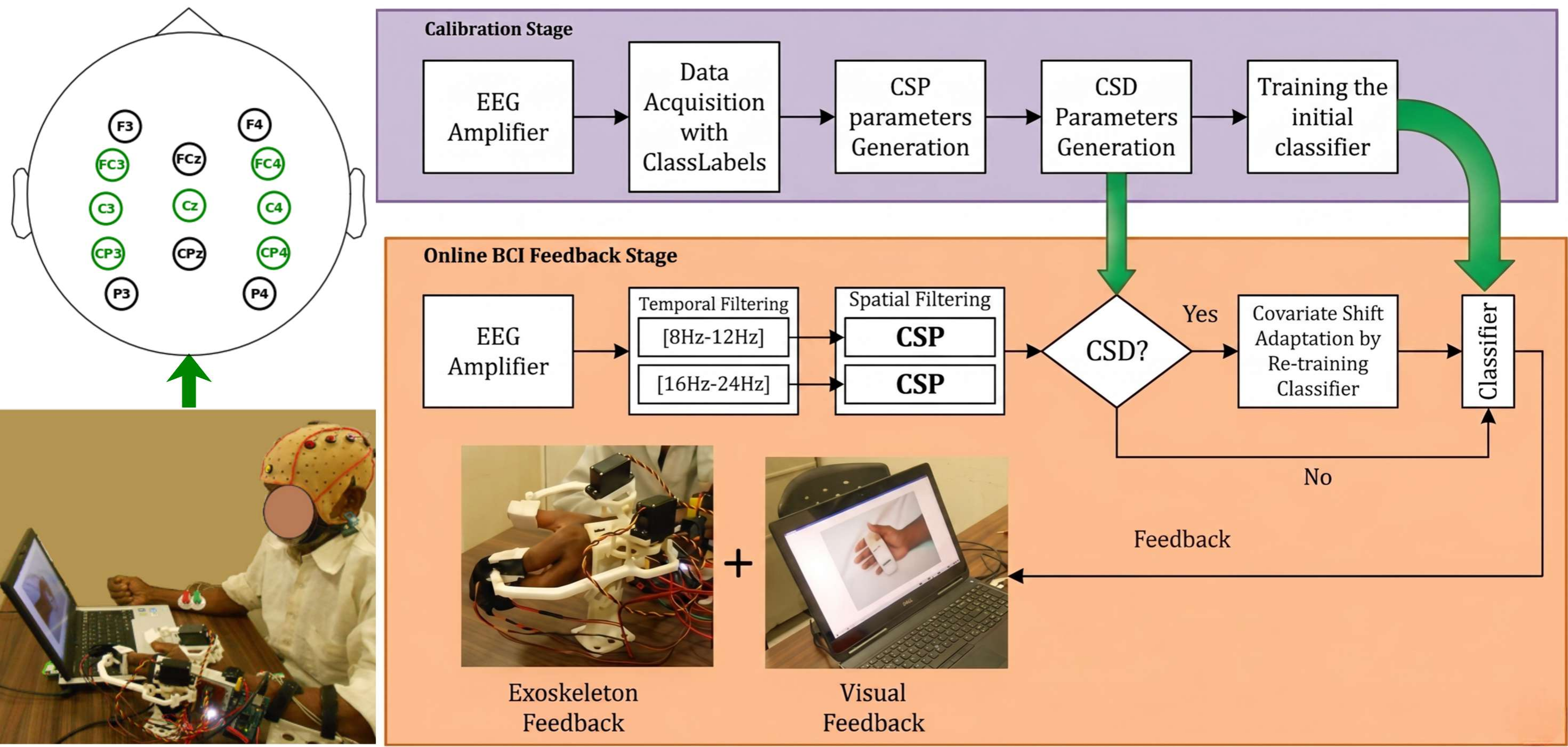}
    \caption{ Experimental protocol for the dataset recorded in previous work~\cite{chowdhury2018eeg}.}
    \label{fig:protocol}
\end{figure}

\begin{figure}[!t]
    \centering
    \includegraphics[width=0.9\linewidth]{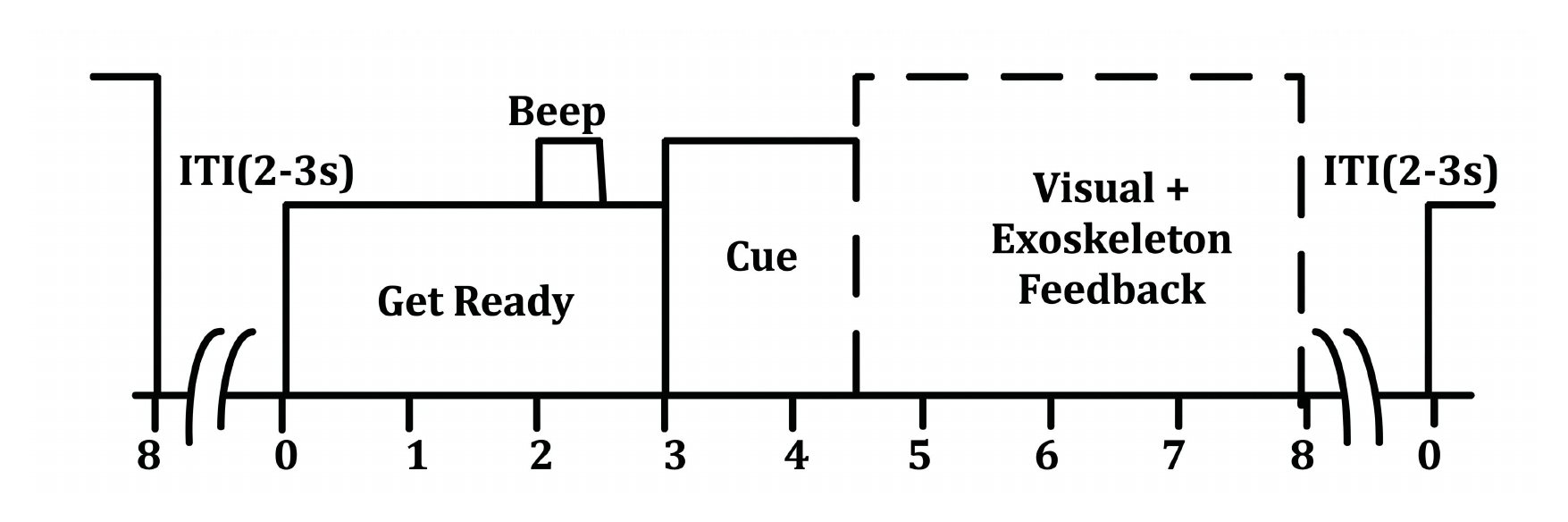}
    \caption{Timing diagram of single-trial dataset recorded in the previous work~\cite{chowdhury2018eeg}.}
    \label{fig:Timing Diagram}
\end{figure}
\begin{figure*}[!t]
    \centering
    \includegraphics[width=0.75\linewidth]{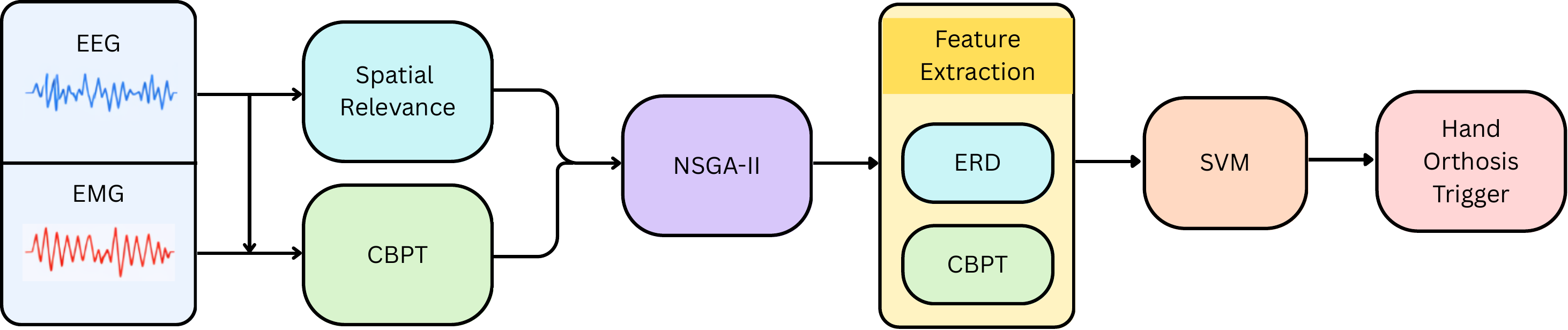}
    \caption{Block Diagram of Proposed EEG-EMG hybrid BCI framework.}
    \label{fig:block_diagram}
\end{figure*}

\section{Materials and Methods}

\subsection{Dataset}

The dataset recorded from eight stroke patients, as described in the previous work~\cite{chowdhury2018eeg}, is used in this work. Figs.~\ref{fig:protocol} and ~\ref{fig:Timing Diagram}, respectively, depict the experimental protocol, which includes data collecting and training/testing timing. The experimental paradigm follows a conventional SMR-based BCI framework comprising two distinct phases: an offline training phase and an online feedback phase, conducted within a single session.

In the first phase, EEG and EMG signals are recorded without providing any feedback to the subject, and the collected data are used to train the classification model. This phase consists of two runs, with each run containing 40 trials and lasting approximately 7-8 minutes. The trials are evenly distributed across the two classes (left-hand and right-hand MI), resulting in 20 trials per class per run. The second phase corresponds to an online BCI setting in which neurofeedback is provided based on the classifier output. This phase consists of a single feedback run of 40 trials, during which the trained model predicts the MI class. A gap of approximately 15-20 minutes is maintained between the training and feedback phases to reduce fatigue and maintain subject attention, which is particularly important in rehabilitation scenarios.

Each trial follows a structured timing protocol in which a visual cue appears approximately 3 seconds after trial onset, indicating the MI task to be performed. The subject performs MI during the 0-2 second interval following the cue. The EEG data are acquired in bipolar mode, resulting in a total of seven bipolar EEG channels derived from the standard 10-20 system. These bipolar derivations include left-hemisphere, midline, and right-hemisphere channels, enabling improved spatial localisation of sensorimotor activity. In addition to EEG, EMG signals are recorded from two forearm muscles, namely the flexor digitorum superficialis of the right and left hands (FDSR and FDSL). The sampling rate and acquisition parameters are consistent across the session and are provided within dataset files.

\subsection{Multi-Objective Optimisation Problem Definition}
In the proposed method, as shown in Fig.~\ref{fig:block_diagram}, the selection of EEG-EMG channel pairs is formulated as a multi-objective optimisation (MOO) problem to identify the most informative cortico-muscular interactions. Unlike conventional approaches that rely on manually predefined channel combinations, the proposed framework adopts a data-driven strategy to automatically determine optimal channel pairs. This is obtained by simultaneously optimising two complementary objectives: the spatial relevance of EEG channels and the cortico-muscular coupling strength between EEG and EMG signals. By integrating these objectives within a unified optimisation framework, the proposed method enables subject-specific adaptation and facilitates the selection of physiologically meaningful and discriminative channel pairs.

\subsubsection{Spatial Relevance}

The first objective considered in this work is the spatial relevance, denoted as $f_{\text{sp}}$, which aims to maximise the contribution of EEG channels located in motor cortex regions. Since MI tasks predominantly activate areas around C3 and C4, channels in close proximity to these locations are expected to be more informative.

To quantify this, a spatial weighting scheme based on a Gaussian kernel is employed~\cite{spatialrelevance}. This formulation assigns higher weights to channels near the reference motor cortex locations while attenuating the influence of distant channels. Let $\mu_j$ denote the spatial weight associated with the $j^{th}$ EEG channel, defined as

\begin{equation}
    \mu_j = \exp\left(-\frac{\|\mathbf{r}_j - \mathbf{r}_{\text{ref}}\|^2}{2\sigma^2}\right)
    \label{eq:gaussian_kernel}
\end{equation}
where $\mathbf{r}_j$ represents the spatial coordinates of the channel and $\mathbf{r}_{\text{ref}}$ denotes the reference motor cortex location. The spatial spread of the kernel is controlled by the parameter $\sigma$. In this study, it is set to $\sigma$ = 0.1 to emphasise channels close to C3 and C4 and enforce a sharp spatial decay.

For a given solution $\mathbf{x}$, representing the selected EEG-EMG pairs, the spatial relevance objective is computed as

\begin{equation}
    f_{\text{sp}}(\mathbf{x}) = \sum_{j \in \mathcal{C}(\mathbf{x})} \mu_j
    \label{eq:spatial_relevance}
\end{equation}

where $\mathcal{C}(\mathbf{x})$ denotes the set of EEG channels involved in the selected pairs. As defined in \eqref{eq:spatial_relevance}, this objective favours the selection of channel pairs concentrated around motor cortex regions, thereby improving physiological interpretability and task relevance.

\subsubsection{Cortico-Muscular Coupling Strength}

The second objective aims to maximise the cortico-muscular coupling strength, denoted as $f_{\text{cm}}$, which captures the interaction between EEG and EMG signals during MI. Strong coupling between cortical activity and muscle responses reflects meaningful neurophysiological patterns and is therefore desirable for accurate classification.
This interaction is quantified using the Pearson correlation coefficient ($\rho_{ij}$) computed between band-limited EEG and EMG signals within the MI interval. 


To obtain a robust estimate, absolute correlation is averaged across all trials, reducing the effect of trial-to-trial variability. The resulting coupling strength is given by

\begin{equation}
    s_{ij} = \frac{1}{T} \sum_{t=1}^{T} |\rho_{ij}(t)|
    \label{eq:pair_score}
\end{equation}

where $T$ denotes the total number of trials. For a candidate solution $\mathbf{x}$, representing the selected EEG-EMG pairs, the coupling objective is defined as

\begin{equation}
    f_{\text{cm}}(\mathbf{x}) = \sum_{(i,j) \in \mathcal{P}(\mathbf{x})} s_{ij}
    \label{eq:cm_objective}
\end{equation}

where $\mathcal{P}(\mathbf{x})$ denotes the set of selected EEG-EMG pairs. As defined in \eqref{eq:cm_objective}, this objective favours the selection of pairs that exhibit strong and consistent coupling, thereby capturing meaningful interactions between cortical activity and muscular responses.

\subsubsection{Pair Selection using NSGA-II}

The EEG-EMG pair selection problem is formulated as a bi-objective optimisation task, where the objective is to simultaneously maximise the spatial relevance and the corticomuscular coupling strength, as defined in Eq. \eqref{eq:spatial_relevance} and \eqref{eq:cm_objective}. The problem is expressed in minimisation form as

\begin{equation} 
    \min_{\mathbf{x}} \left[-f_{\text{sp}}(\mathbf{x}), \, -f_{\text{cm}}(\mathbf{x})\right] 
    \label{eq:moo_problem} 
    \end{equation} 
    
subject to the cardinality constraint $\sum_{i=1}^{N} x_i \leq K$,

where $N$ denotes the total number of candidate EEG-EMG pairs, $K$ is the maximum number of pairs allowed, and $\mathbf{x} \in \{0,1\}^N$ represents the selection vector. This binary encoding is well-suited for the pair selection problem, as it directly models the inclusion or exclusion of individual EEG-EMG pairs in a compact and interpretable manner. 

To improve computational efficiency, both objective components are precomputed prior to optimisation. The spatial relevance weights are obtained using the Gaussian kernel described in Eq.~\eqref{eq:gaussian_kernel}, while the corticomuscular coupling strength for each EEG-EMG pair is computed as the average absolute correlation across trials. 
During optimisation, objective values for a candidate solution are then evaluated by aggregating these precomputed quantities over the selected pairs.

The optimisation problem is solved using the NSGA-II algorithm~\cite{NSGA2}, implemented via the \textit{pymoo} framework. Each individual in the population represents a candidate subset of EEG-EMG pairs encoded as a binary vector. The algorithm employs non-dominated sorting and crowding distance mechanisms to evolve a diverse set of Pareto-optimal solutions. Standard genetic operators, including binary crossover and mutation, are applied internally within the optimisation framework to explore the search space and maintain diversity among candidate solutions.

From the obtained Pareto-optimal set, a single solution is selected based on a combined objective score that balances spatial relevance and coupling strength. To ensure a compact and interpretable feature set, a maximum of $K = 4$ EEG-EMG pairs is retained. If a candidate solution contains more than $K$ pairs, only the top-ranked pairs based on coupling strength are preserved. This selection strategy ensures that the final subset of pairs is both physiologically meaningful and discriminative for MI classification.

\subsection{Feature Extraction}
After selecting the optimal EEG-EMG channel pairs, the signals are first subjected to preprocessing to isolate relevant frequency components and reduce noise. EEG signals are band-pass filtered within the sensorimotor rhythm ranges, namely the $\mu$ (8-12 Hz) and $\beta$ (16-24 Hz) bands, while EMG signals are filtered in the 30-50 Hz band. In addition, a 50 Hz notch filter is applied to eliminate power-line interference. To enhance robustness, an ERD-based band selection strategy is employed, wherein the frequency band exhibiting lower inter-trial variability is selected for subsequent analysis. For the selected optimal pairs, CBPT and ERD features are extracted to capture corticomuscular interactions and task-related neural activity.

For each selected EEG–EMG pair, the band-limited EEG and EMG signals are transformed into power time series and smoothed using moving-average filters. The Pearson correlation coefficient is then computed between the EEG and EMG power signals within each temporal window, and its absolute value is used as the CBPT measure. To suppress weak and potentially spurious interactions, correlation values with absolute magnitudes below 0.1 are set to 0, while those greater than or equal to 0.1 are retained.

In addition to CBPT features~\cite{chowdhury2018eeg}, ERD-based spectral features are extracted from the bipolar EEG channels at C3, Cz, and C4. Logarithmic band power is computed for each channel, and a hemispheric asymmetry feature is obtained as the difference between C4 and C3, capturing task-related modulation of sensorimotor rhythms during MI. Feature extraction is performed using a sliding window approach over the 0-2 s post-cue interval. Each window has a duration of 1 s with a step size of 128 ms, yielding multiple candidate feature sets across time.
\subsection{Experimental Setup and Classification}
For classification, a support vector machine (SVM) with a radial basis function (RBF) kernel is used~\cite{7318410, 6999180}. The SVM hyperparameters, including the regularisation parameter $C$ and kernel parameter $\gamma$, are optimised using grid search. For each temporal window identified during feature extraction, the model is trained on features from the training data and evaluated using stratified 10-fold cross-validation. The window yielding the highest cross-validation accuracy is selected and SVM is retrained on the full training set using that window before being evaluated on feedback data.

For each trial, the feature vector consists of CBPT features from the $K$ selected EEG-EMG pairs and four ERD-based features (log-power of C3, Cz, and C4, along with the C4-C3 asymmetry), yielding a total of $(K+4)$ features per window. In this work, $K = 4$ pairs are retained following the MOO-based selection, resulting in an 8-dimensional feature vector.

With a population size of 100 across 150 generations and crossover and mutation probabilities set to 0.9 and 0.07, respectively, the algorithm is executed in the NSGA-II setup. Also note that in the original dataset experiment, the classifier was built on CSP-based features and retrained during online neurofeedback using the covariate-shift detection (CSD) technique~\cite{chowdhury2018eeg}; the present work replaces this pipeline with the proposed hybrid EEG-EMG framework described above.

\begin{table}[]
\caption{Selected EEG-EMG channel pairs and corresponding classification accuracy for each patient. Pairs are obtained using the proposed NSGA-II-based optimisation framework.}
\label{tab:selected_pairs}
\begin{tabular}{@{}ccl@{}}
\toprule
\textbf{Patient} & \textbf{Acc (\%)} & \textbf{Selected Channel Pairs} \\ \midrule
P1 & 100.0 & FC3-FDSR, C3-FDSR, CP3-FDSR, Cz-FDSR \\
P2 & 85.0 & CP3-FDSR, Cz-FDSR, C4-FDSL, CP4-FDSL \\
P3 & 90.0 & CP3-FDSR, FC4-FDSL, C4-FDSL, CP4-FDSL \\
P4 & 92.5 & CP3-FDSR, FC4-FDSL, C4-FDSL, CP4-FDSL \\
P5 & 100.0 & CP3-FDSR, Cz-FDSR, Cz-FDSL, CP4-FDSL \\
P6 & 85.0 & C3-FDSR, CP3-FDSR, Cz-FDSR, Cz-FDSL \\
P7 & 82.5 & CP3-FDSR, FC4-FDSL, C4-FDSL, CP4-FDSL \\
P8 & 82.5 & FC3-FDSR, C3-FDSR, CP3-FDSR, Cz-FDSL \\ \bottomrule
\end{tabular}
\end{table}

        

\begin{figure}[!t]
    \centering
    \includegraphics[width=0.7\linewidth]{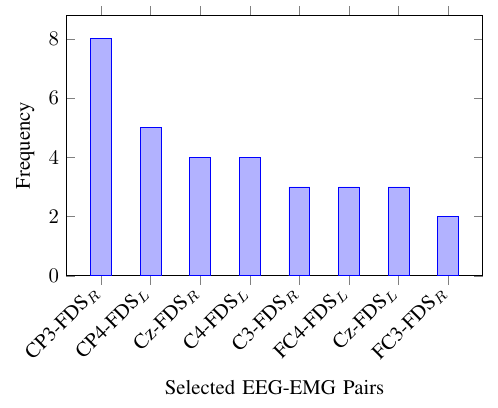}
    \caption{Frequency of selected channel pairs across patients.}
    \label{fig:pair_frequency}
\end{figure}

\section{Results}
To evaluate the effectiveness of the proposed pair selection strategy, the performance was compared with a baseline approach that utilises CBPT features extracted from fixed EEG-EMG pairs (involving C3, Cz, and C4) without optimisation~\cite{chowdhury2018eeg}. The classification accuracy for each subject using the proposed method is presented in Table~\ref{tab:selected_pairs}. The baseline approach achieves an average accuracy of 84.53\% as reported in~\cite{chowdhury2018eeg}, whereas the proposed framework improves performance to 89.6\%, yielding an absolute gain of approximately 5\%. This improvement highlights the benefit of automatically selecting informative EEG-EMG channel pairs over relying on predefined configurations.



\begin{figure}
    \centering
    \includegraphics[width=0.75\linewidth]{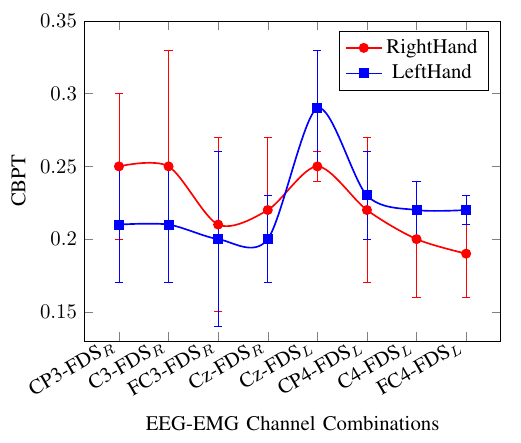}
\caption{Average CBPT feature distribution across channel pairs arranged to reflect spatial progression from left to right sensorimotor cortex.}
\label{fig:cbpt_all_pairs}
\end{figure}
An analysis of the selected channel pairs reveals consistent patterns across patients, with a strong emphasis on EEG channels located around the sensorimotor cortex. Channels such as C3, CP3, Cz, and C4 are frequently selected, confirming their relevance to MI tasks. In addition, a clear contralateral pairing trend is observed, where EEG channels from the left hemisphere are predominantly associated with right-hand EMG signals (e.g., CP3-FDSR), and vice versa (e.g., C4-FDSL). This behaviour aligns with established neurophysiological principles of motor control.

The consistency of the selected pairs is further illustrated in Fig.~\ref{fig:pair_frequency}, which shows the frequency of occurrence of each EEG-EMG pair across patients. CP3-FDSR is selected in all 8 patients, making it the most dominant pair, followed by CP4-FDSL (5 patients). Cz-FDSR and C4-FDSL each appear in 4 patients, while C3-FDSR, FC4-FDSL, and Cz-FDSL are selected in 3 patients, and FC3-FDSR appears in 2 patients. The repeated selection of these pairs across subjects indicates that they consistently capture meaningful corticomuscular interactions, with dominant pairs such as CP3-FDSR and CP4-FDSL reflecting strong coupling between sensorimotor cortex regions and corresponding muscles.

To further examine the discriminative capability of the selected pairs, the average correlation values for left and right hand MI tasks are reported in Fig.~\ref{fig:cbpt_all_pairs}. Several channel-muscle pairs exhibit relatively high and stable correlation values, indicating strong corticomuscular coupling. For instance, CP3-FDSR shows a correlation of $0.21 \pm 0.04$ for left-hand imagery and $0.25 \pm 0.05$ for right-hand imagery, while C3-FDSR follows a similar trend with $0.21 \pm 0.04$ and $0.25 \pm 0.08$, respectively. This indicates stronger coupling during right-hand tasks for left hemisphere EEG channels, consistent with contralateral motor control.

In contrast, Cz-FDSL demonstrates higher correlation for left hand imagery ($0.29 \pm 0.04$) compared to right hand imagery ($0.25 \pm 0.01$), highlighting task-dependent modulation of corticomuscular interaction. Similarly, CP4-FDSL shows slightly higher coupling for left hand ($0.23 \pm 0.03$) compared to right hand ($0.22 \pm 0.05$), while C4-FDSL ($0.22 \pm 0.02$, $0.20 \pm 0.04$) and FC4-FDSL ($0.22 \pm 0.01$, $0.19 \pm 0.03$) also exhibit stronger interaction during left hand tasks. These observations reflect increased coupling in right-hemisphere channels for left-hand MI.

Higher correlation values indicate stronger functional coupling between cortical activity and muscle activation, while lower variability reflects more stable and reliable interaction patterns across trials. These findings are consistent with pairs identified by the optimisation framework. Frequently selected pairs such as CP3-FDSR and CP4-FDSL exhibit stable and relatively high correlation values, confirming their importance in capturing corticomuscular dynamics. Other pairs, including FC3-FDSR ($0.20 \pm 0.06$, $0.21 \pm 0.06$) and Cz-FDSR ($0.20 \pm 0.03$, $0.22 \pm 0.05$), also contribute to the overall interaction, indicating that multiple channel combinations are involved in representing MI.

The CBPT distribution across EEG-EMG channel pairs, arranged to reflect the spatial progression from the left to the right sensorimotor cortex, is shown in Fig.~\ref{fig:cbpt_all_pairs}. A clear spatial trend is observed, where left-hemisphere pairs (CP3-FDSR, C3-FDSR) exhibit higher CBPT values for right-hand imagery, while right-hemisphere pairs (Cz-FDSL, CP4-FDSL) show stronger coupling for left-hand imagery. This indicates a transition (flipped) in CBPT values across hemispheres. In particular, Cz-FDSL shows the highest CBPT value for the left-hand task, whereas CP3-FDSR exhibits stronger coupling for the right-hand task. This 
reflects the contralateral organisation of motor control and demonstrates that CBPT features capture task-dependent corticomuscular interactions.


\section{Discussion}
The results demonstrate that proposed approach improves MI classification performance compared to the baseline method, highlighting the importance of selecting informative EEG-EMG channel pairs rather than relying on fixed configurations. A key observation is the consistent selection of EEG channels around the sensorimotor cortex, such as C3, CP3, Cz, and C4, which are closely associated with motor-related activity. Furthermore, the dominance of contralateral EEG-EMG pairings, where left-hemisphere EEG channels correspond to right hand EMG activity and vice versa, aligns with established neurophysiological principles of motor control.

The frequency analysis reveals that certain channel pairs, such as CP3-FDSR and C4-FDSL, are consistently selected across multiple patients, indicating their robustness in capturing corticomuscular interactions. At the same time, variations in selected pairs across subjects reflect inter-subject variability, emphasising the need for subject-specific channel selection. This behaviour highlights the advantage of the NSGA-II-based adaptive pair selection, which enables the identification of both consistently strong and task-specific corticomuscular interactions.

Another important finding is that high classification performance can be achieved with a limited number of channel pairs, indicating that a compact feature representation is sufficient to capture relevant corticomuscular information. This reduces computational complexity and improves the feasibility of real-time BCI systems. Compared to the CBPT-based framework in~\cite{chowdhury2018eeg}, which relies on fixed EEG channels, the proposed method introduces an optimisation-driven pair selection strategy that adapts to subject-specific corticomuscular patterns. These results demonstrate that integrating data-driven pair selection with CBPT features improves both classification performance and physiological interpretability.
\section{Conclusion}
This work presents a data-driven EEG-EMG channel-pair selection framework for hybrid BCI systems, where channel-pair selection is formulated as an MOO problem. By jointly considering spatial relevance and corticomuscular coupling, the proposed method identifies informative, physiologically meaningful channel pairs, thereby improving classification performance compared to fixed channel pair configurations. The results show that adaptive pair selection enhances the effectiveness of CBPT features while maintaining a compact and interpretable representation. The observed consistency of selected pairs across subjects, along with subject-specific variations, highlights the approach's ability to capture both common and individual corticomuscular patterns. In future, the proposed framework can be evaluated on larger, diverse datasets involving multimodal physiological signals to assess its robustness and generalisability. 
As recent studies have highlighted the importance of the gamma band in motor activation and movement-related tasks \cite{kacker2025motor}, this work would can be extended to explore the importance.
Further improvements can be achieved by exploring more efficient optimisation strategies for real-time deployment in practical BCI applications. The proposed framework, therefore, provides a scalable, physiologically grounded approach to improving hybrid BCI systems.
All source code and dataset details are available at \url{https://github.com/HAIx-Lab/EEG-EMG-Pair-Selection-Framework} to ensure reproducibility of this work. 

\noindent{\textbf{Acknowledgement:}} This study was supported by the IIT Gandhinagar startup grant (IP/IITGN/CSE/YM/2324/05). 



\bibliographystyle{IEEEtran}

\bibliography{main}

@article{chowdhury2018eeg,
  author  = {Chowdhury, Anirban and Raza, Haider and Meena, Yogesh Kumar and Dutta, Ashish and Prasad, Girijesh},
  title   = {An EEG--EMG correlation-based brain--computer interface for hand orthosis supported neuro-rehabilitation},
  journal = {Journal of Neuroscience Methods},
  volume  = {312},
  pages   = {1--11},
  year    = {2019},
  doi     = {10.1016/j.jneumeth.2018.11.010}
}

@ARTICLE{spatialrelevance,
  author={Handiru, Vikram Shenoy and Prasad, Vinod A.},
  journal={IEEE Transactions on Human-Machine Systems}, 
  title={Optimized Bi-Objective EEG Channel Selection and Cross-Subject Generalization With Brain–Computer Interfaces}, 
  year={2016},
  volume={46},
  number={6},
  pages={777-786},
  doi={10.1109/THMS.2016.2573827}}

@article{NSGA2,
author = {Deb, K. and Pratap, A. and Agarwal, S. and Meyarivan, T.},
title = {A fast and elitist multiobjective genetic algorithm: NSGA-II},
year = {2002},
issue_date = {April 2002},
publisher = {IEEE Press},
volume = {6},
number = {2},
issn = {1089-778X},

journal = {Trans. Evol. Comp},
month = apr,
pages = {182–197},
numpages = {16}
}

@article{wolpaw2002brain,
  author  = {McFarland, Dennis J. and Wolpaw, Jonathan R.},
  title   = {Brain--computer interfaces for communication and control},
  journal = {Communications of the ACM},
  volume  = {54},
  number  = {5},
  pages   = {60--66},
  year    = {2011}
}

@ARTICLE{pfurtscheller2001motor,
  author={Pfurtscheller, G. and Neuper, C.},
  journal={Proceedings of the IEEE}, 
  title={Motor imagery and direct brain-computer communication}, 
  year={2001},
  volume={89},
  number={7},
  pages={1123-1134}}

@article{pfurtscheller1999event,
  author  = {Pfurtscheller, G. and Lopes da Silva, F. H.},
  title   = {Event-related EEG/MEG synchronization and desynchronization: basic principles},
  journal = {Clinical Neurophysiology},
  volume  = {110},
  number  = {11},
  pages   = {1842--1857},
  year    = {1999},

}

@article{lotte2018review, 

year = {2018}, 
month = {apr}, 
publisher = {IOP Publishing}, 
volume = {15}, 
number = {3}, 
journal = {Journal of Neural Engineering},
pages = {031005}, 
author = {Lotte, F and Bougrain, L and Cichocki, A and Clerc, M and Congedo, M and Rakotomamonjy, A and Yger, F}, 
title = {A review of classification algorithms for EEG-based braincomputer interfaces (BCIs)} 
}

@article{conway1995synchronization,
author = {Conway, B A and Halliday, D M and Farmer, S F and Shahani, U and Maas, P and Weir, A I and Rosenberg, J R},
title = {Synchronization between motor cortex and spinal motoneuronal pool during the performance of a maintained motor task in man.},
journal = {The Journal of Physiology},
volume = {489},
number = {3},
pages = {917-924},

year = {1995}
}

@article{blankertz2008optimizing,
author = {Blankertz, Benjamin and Tomioka, R. and Lemm, S. and Kawanabe, Motoaki and Müller, Klaus-Robert},
year = {2008},
month = {01},
pages = {41-56},
title = {Optimizing spatial filters for robust EEG single-trial analysis},
volume = {25},
journal = {IEEE Signal. Proc. Mag.}
}

@article{halliday1998coherence,
title = {A framework for the analysis of mixed time series/point process data:Theory and application to the study of physiological tremor, single motor unit discharges and electromyograms},
journal = {Progress in Biophysics and Molecular Biology},
volume = {64},
pages = {237-278},
year = {1995},
issn = {0079-6107},

author = {D.M. Halliday et al.}
}

@article{pfurtscheller2003cmc,
  title={Corticomuscular coherence in motor control},
  author={Pfurtscheller, Gert and others},
  journal={Journal of Clinical Neurophysiology},
  year={2003}
}

@ARTICLE{tuncel2010cmc,
	author = {Hashimoto, Yasunari and Ushiba, Junichi and Kimura, Akio and Liu, Meigen and Tomita, Yutaka},
	title = {Correlation between EEG-EMG coherence during isometric contraction and its imaginary execution},
	year = {2010},
	journal = {Acta Neurobiologiae Experimentalis},
	volume = {70},
	number = {1},
	pages = {76 – 85},
	type = {Article},
	publication_stage = {Final},
	source = {Scopus}
}

@article{allison2011hybrid,
  author  = {Allison, Brendan Z. and Brunner, Clemens and Kaiser, Viktoria and M{\"u}ller-Putz, Gernot R. and Neuper, Christa and Pfurtscheller, Gert},
  title   = {Toward a hybrid brain--computer interface based on imagined movement and visual attention},
  journal = {Journal of Neural Engineering},
  volume  = {7},
  number  = {2},
  pages   = {026007},

}

@ARTICLE{muller2011hybrid,
    
AUTHOR={Pfurtscheller, Gert  and Allison, Brendan Z. and Bauernfeind, Günther  and Brunner, Clemens  and Solis Escalante, Teodoro  and Scherer, Reinhold  and Zander, Thorsten O. and Mueller-Putz, Gernot  and Neuper, Christa  and Birbaumer, Niels },
           
TITLE={The hybrid BCI},
          
JOURNAL={Frontiers in Neuroscience},
          
VOLUME={Volume 4 - 2010},
  
YEAR={2010},
  
  
  
ISSN={1662-453X},
}

@article{kacker2025motor,
  title={Motor activity in gamma and high gamma bands recorded with a Stentrode from the human motor cortex in two people with ALS},
  author={Kacker, Kriti and Chetty, Nikole and Feldman, Ariel K and Bennett, James and Yoo, Peter E and Fry, Adam and Lacomis, David and Harel, Noam Y and Nogueira, Raul G and Majidi, Shahram and others},
  journal={Journal of Neural Engineering},
  volume={22},
  number={2},
  pages={026036},
  year={2025},
  publisher={IOP Publishing}
}

@article{muller2015brain,
author = {Pfurtscheller, Gert and Müller-Putz, Gernot and Scherer, Reinhold and Neuper, Christa},
year = {2008},
month = {11},
pages = {58 - 65},
title = {Rehabilitation with Brain-Computer Interface Systems},
volume = {41},
journal = {Computer},
doi = {10.1109/MC.2008.432}
}

@article{liu2013emg,
  title={EMG-based control systems in rehabilitation},
  author={Liu, Jie and Zhou, Ping},
  journal={Biomedical Engineering},
  year={2013}
}

@INPROCEEDINGS{makowski2015emg,
  author={Alsayegh, O.A.},
  booktitle={2000 IEEE International Conference on Multimedia and Expo. ICME2000. Proceedings. Latest Advances in the Fast Changing World of Multimedia (Cat. No.00TH8532)}, 
  title={EMG-based human-machine interface system}, 
  year={2000},
  volume={2},
  number={},
  pages={925-928 vol.2},
 }

@article{cincotti2012neurofeedback,
  author  = {Renton, T. and Tibbles, A. and Topolovec-Vranic, J.},
  title   = {Neurofeedback as a form of cognitive rehabilitation therapy following stroke: A systematic review},
  journal = {PLoS ONE},
  volume  = {12},
  number  = {5},
  pages   = {e0177290},
  year    = {2017},
  doi     = {10.1371/journal.pone.0177290}
}

@article{baig2019survey,
  author  = {Baig, M. Z. and Aslam, N. and Shum, H. P. H.},
  title   = {Filtering techniques for channel selection in motor imagery EEG applications: a survey},
  journal = {Artificial Intelligence Review},
  volume  = {53},
  pages   = {1207--1232},
  year    = {2020},
  doi     = {10.1007/s10462-019-09694-8}
}

@Article{faye2022review,
AUTHOR = {Abdullah and Faye, Ibrahima and Islam, Md Rafiqul},
TITLE = {EEG Channel Selection Techniques in Motor Imagery Applications: A Review and New Perspectives},
JOURNAL = {Bioengineering},
VOLUME = {9},
YEAR = {2022},
NUMBER = {12},
ARTICLE-NUMBER = {726},
PubMedID = {36550932},
ISSN = {2306-5354}
}

@article{Lee2025HybridEEG,
  author    = {Lee, HT and Shim, M and Liu, X and et al.},
  title     = {A review of hybrid EEG-based multimodal human--computer interfaces using deep learning: applications, advances, and challenges},
  journal   = {Biomedical Engineering Letters},
  year      = {2025},
  volume    = {15},
  pages     = {587--618},

  issn      = {2093-9868}
}

@INPROCEEDINGS{6999180,
  author={O'Doherty, Darren and Meena, Yogesh Kumar and Raza, Haider and Cecotti, Hubert and Prasad, Girijesh},
  booktitle={2014 IEEE International Conference on Bioinformatics and Biomedicine (BIBM)}, 
  title={Exploring gaze-motor imagery hybrid brain-computer interface design}, 
  year={2014},
  volume={},
  number={},
  pages={335-339},
  doi={10.1109/BIBM.2014.6999180}}

@inproceedings{meena2015simultaneous,
title={Simultaneous gaze and motor imagery hybrid bci increases single-trial detection performance: a compatible incompatible study},
author={Meena, Yogesh and Prasad, Girijesh and Cecotti, Hubert and Wong-Lin, KongFatt},
booktitle={9th IEEE-EMBS International Summer School on Biomedical Signal Processing},
year={2015},
organization={IEEE Engineering in Medicine and Biology Society}
}

@INPROCEEDINGS{7318410,
  author={Meena, Yogesh Kumar and Cecotti, Hubert and Wong-Lin, KongFatt and Prasad, Girijesh},
  booktitle={2015 37th Annual International Conference of the IEEE Engineering in Medicine and Biology Society (EMBC)}, 
  title={Towards increasing the number of commands in a hybrid brain-computer interface with combination of gaze and motor imagery}, 
  year={2015},
  volume={},
  number={},
  pages={506-509},
  doi={10.1109/EMBC.2015.7318410}}

\end{document}